\documentclass[proof]{WileyASNA-v1}

\articletype{Article Type}%

\received{26 April 2018}
\revised{6 June 2018}
\accepted{6 June 2018}

\newcommand{\beq}{\begin{equation}}
\newcommand{\eeq}{\end{equation}}
\newcommand{\bea}{\begin{eqnarray}}
\newcommand{\eea}{\end{eqnarray}}

\newcommand{\ce}{{\cal{E}}} 
\newcommand{\cl}{{\cal{L}}} 

\newcommand{\BB}{{\cal{B}}}

\begin{document}

\title{Orbital widening due to radiation reaction around a magnetized black hole 
}

\author{Arman Tursunov*}

\author{Martin Kolo{\v s}}

\author{Zden{\v e}k Stuchl{\'i}k}


\address{
\orgdiv{Institute of Physics and Research Centre of Theoretical Physics and Astrophysics}, \orgname{Faculty of Philosophy and Science, \\ Silesian University in Opava}, \orgaddress{Bezru{\v c}ovo n{\'a}m.13, CZ-74601 Opava, \country{Czech Republic} }
}

\corres{*Arman Tursunov, Bezru{\v c}ovo n{\'a}m.13, CZ-74601 Opava, Czech Republic. \email{arman.tursunov@fpf.slu.cz}}


\abstract{ Radiation reaction acting on a charged particle moving at a stable circular orbit of a magnetized black hole can lead to the shift of the orbital radius outwards from the black hole. The effect causes increase of the energy and angular momentum of the particle measured by an observer at rest at infinity.
In this note we show that "widening" of such orbits is independent of the field configuration, however, it appears only in the cases with the external Lorentz force acting outwards from the black hole. This condition corresponds to $q L B > 0$, where $q$ and $L$ are the charge and angular momentum of the particle and $B$ is intensity of the external magnetic field. As examples of the orbital widening we consider two scenarios with an external homogeneous magnetic field and a magnetic dipole field generated by a current loop around a Schwarzschild black hole. We show that the orbital widening is accompanied by quasi-harmonic oscillations of the particle which are considerably large in the magnetic dipole fields. {We also estimate the timescales of orbital widening from which it follows that the effect can be relevant in the vicinity of stellar mass black holes.}
}

\keywords{radiation mechanism: non-thermal, black hole physics, relativity, magnetic fields}



\maketitle


\section{Introduction}\label{introduc}

It has been pointed out in \cite{Tur-etal:2018:APJ:} that one of the consequences of the synchrotron self-force in the vicinity of a black hole, and in presence of an external magnetic field, is the evolution of the circular orbits of charged particles during the radiation process and shifting of the radius of the circular orbits outwards from the black hole. This can be realized for repulsive Lorentz force only, while in attractive case, the particle spirals down to the black hole and no closed circular orbits can be formed.
In this paper we investigate the effect of widening of the stable circular orbit of charged particle undergoing radiation reaction force around a Schwarzschild black hole immersed in a uniform, or a dipole, magnetic field. The effect is considerable, if the radiation reaction cannot be neglected and appears, as we show, due to the diversity in the dependencies of the kinetic and potential energies of a charged particle on the radius of its orbit. Although the kinetic energy of the particle decreases due to the synchrotron radiation, its potential energy increases faster, due to the widening of the orbit. 
In fact, the self-force decreases the components of four-velocity of the particle with a rate proportional to the derivative of the field. If the stability of the orbit is guaranteed by the location of the particle at the minimum of the effective potential, the energy of the particle measured at infinity can start to increase.

Moreover, the orbital widening leads to appearance of quasi-harmonic oscillations of charged particles with amplitudes increasing with increasing energy of the particle.
{Hereafter in the present paper, we use the geometrized system of units with $G=1=c$, however, we insert physical constants back in Conclusions while discussing the astrophysical relevance of the presented effect.} Greek indexes are running from $0$ to $3$.

\section{Radiation reaction force}\label{sec1}

In many astrophysically relevant scenarios one cannot neglect the effects of radiation reaction due to the synchrotron radiation of charges in the vicinity of black holes, which are believed to be immersed into an external magnetic field.
It is important to identify the conditions of stability of circular orbits in such regime. Dynamics of charged particles undergoing radiation reaction force in the flat spacetime is 
governed by the Lorenz-Dirac equation, while dynamics in curved spacetimes is described by the DeWitt-Brehme equation \citep{DeW-Bre:1960:AnnPhys:}, completed by \cite{Hobbs:1968:AnnPhys:}. The resulting equation of motion of a charged particle in a curved spacetime reads \citep{Poisson:2004:LRR:}
\bea 
&& \frac{D u^\mu}{d \tau} = \frac{q}{m} F^{\mu}_{\,\,\,\nu} u^{\nu} 
+ \frac{2 q^2}{3 m} \left( \frac{D^2 u^\mu}{d\tau^2} + u^\mu u_\nu \frac{D^2 u^\nu}{d\tau^2} \right) 
\nonumber \\
&& + \frac{q^2}{3 m} \left(R^{\mu}_{\,\,\,\lambda} u^{\lambda} 
+ R^{\nu}_{\,\,\,\lambda} u_{\nu} u^{\lambda} u^{\mu} \right) 
+ \frac{2 q^2}{m} ~f^{\mu \nu}_{\rm \, tail} \,\, u_\nu, 
\label{eqmoDWBH}  
\eea 
where in the last term of Eq.(\ref{eqmoDWBH}) the tail integral reads
\beq
f^{\mu \nu}_{\rm \, tail}  = 
\int_{-\infty}^{\tau-0^+}     
D^{[\mu} G^{\nu]}_{ + \lambda'} \bigl(z(\tau),z(\tau')\bigr)   
u^{\lambda'} \, d\tau' .
\eeq
{Here $u^\mu$ is the four-velocity of the particle with charge $q$ and mass $m$. The tensor of electromagnetic field is $F_{\mu\nu} = A_{\nu,\mu} - A_{\mu,\nu}$, where $A_{\mu}$ is the four-vector potential of external electromagnetic field. $R^{\mu}_{\,\,\,\nu}$ is the Ricci tensor, $G^{\mu}_{ + \lambda'}$ is the retarded 
Green's function, and the integration %
is taken along the %
worldline of the particle $z$, %
i.e., $u^\mu (\tau) = d z^\mu(\tau) /d \tau $.
}
The tail integral is calculated over the history of the charged particle, where primes indicate its prior positions. Other quantities in (\ref{eqmoDWBH}) are evaluated at the given position of the particle $z(\tau)$.
The Ricci term is irrelevant, as it vanishes in the vacuum metrics, while the tail term can be neglected for elementary particles, as shown in \cite{Tur-etal:2018:APJ:} and references therein. In particular, for electrons the ratio of the "tail" force and the Newtonian "gravitational" force at the horizon of a black hole of $10$ solar masses is of the order of $10^{-19}$. Since there are no any convincing evidences of the deviations of metrics of astrophysical black holes from vacuum Kerr solution, the motion of electrons and protons in the vicinity of astrophysical black holes can be well described by the covariant form of the Lorentz-Dirac equation. The Lorentz-Dirac equation contains the Schott term -- the third order derivative of coordinate,
which leads to the appearance of pre-accelerating solutions in the absence of external forces. However, one can effectively reduce the order of the equation by substituting the third order terms by derivatives of the external force. This is identical to imposing 
Dirac's asymptotic condition $ \frac{D u^{\mu}}{d \tau}|_{\tau \rightarrow \infty} = 0$ \citep{Spohn:2000:EPL:}. Then, the resulting equation of motion reads
\bea \label{curradforce}
&& \frac{D u^\mu}{d \tau} = \frac{q}{m} F^{\mu}_{\,\,\,\nu} u^{\nu} \nonumber \\
&& + \,  \frac{2 q^2}{3 m} \left(F^{\alpha}_{\,\,\,\beta ; \mu} u^\beta u^\mu + \frac{q}{m} \left( F^{\alpha}_{\,\,\,\beta} 
F^{\beta}_{\,\,\,\mu} +  F_{\mu\nu} F^{\nu}_{\,\,\,\sigma} u^\sigma u^\alpha \right) u^\mu \right),
\eea
where semicolon denotes the covariant coordinate derivative. The equation (\ref{curradforce}), which is usually called as Landau-Lifshitz equation \citep{Lan-Lif:1975:CTF:}, is a habitual second order differential equation which satisfies the principle of inertia and does not contain runaway solutions. Below we use this form of equation for the description of the orbital widening of charged particles in magnetized black hole vicinity. 

{More details on the problem of electromagnetic radiation of point charged particles and related self-force can be found in e.g. \cite{Smi-Wil:1980:PRD:,Sokolov-etal:1983:SovPJ:,Zerilli:2000:NCBS:,Galtsov:2002:PRD:,Poisson:2004:LRR:,Gal-Spi:2006:GRC:,Pri-Bel-Nic:2013:AmerJP:,Barack:2014:book:}.}

\section{Radiative widening of circular orbits }\label{wideningg}

Computational ways of integration of dynamical equation (\ref{curradforce}) and corresponding analyses of trajectories were presented in \cite{Tur-etal:2018:APJ:}. In particular it was shown that for equatorial motion of a charged particle depending on the orientation of external Lorentz force, the final fate of the particle is either collapse to the black hole or stable circular orbit, i.e., the radiation reaction force leads to decay of any oscillations around equilibrium radius.
Here, we concentrate attention on the question what happens once the radiation reaction is taken into account for the charged particle at the stable circular orbit around black hole in the presence of external magnetic field?

Let us consider a purely circular motion of a charged particle revolving around black hole at the equatorial plane, in the presence of a uniform, or a dipole magnetic field. We will specify the components of the fields in the next two subsections. The metric of the Schwarzschild black hole spacetime reads
\beq
 d s^2 = - \left(1 - \frac{2 M}{r}\right) d t^2 + \left(1 - \frac{2 M}{r}\right)^{-1} d r^2 + r^2 d\Omega^2. \label{SCHmetric}
\eeq
The four-velocity of the charged particle satisfies the following equations
\beq \label{norconflat}
u_\alpha u^{\alpha} = -1, \quad u_\alpha \dot{u}^{\alpha} = 0, \quad u_\alpha \ddot{u}^{\alpha} = - \dot{u}_{\alpha} \dot{u}^{\alpha}.
\eeq
In case when the radiation reaction force can be neglected, the symmetry of background spacetime allows one to determine two integrals of motion
\beq
{\cal E} = \left(1-\frac{2M}{r}\right) u^t,  \quad {\cal L} = r^2 \left(u^{\phi} + \frac{q}{m} A_{\phi} \right),
\eeq
which are the specific energy and specific angular momentum measured by an observer at rest at infinity. In general, when the radiation reaction cannot be neglected,  the quantities ${\cal E}$ and ${\cal L}$ are not conserved.

{Oscillatory motion of charged particles and related circular orbits in magnetized Schwarzschild black hole spacetime were studied by \cite{Fro-Sho:2010:PHYSR4:,Kol-Stu-Tur:2015:CQG:}, in magnetized Kerr spacetime by \cite{Shi-Kim-Chi:2014:PRD:,Tur-Stu-Kol:2016:PHYSR4:,Kol-Tur-Stu:2017:EPJC:}, in magnetized "Ernst" metric by  \cite{Lim:2015:PRD:}, as well as in other papers, e.g. \cite{Gal-Pet:1978:SovJETP:,Ali-Gal:1989A:SovPU:,Abd-Ahm-Kag:2008:GRG:,Abd-Ahm:2009:APSS:,Rah-Abd-Ahm:2011:APSS:,Tur-etal:2013:PRD:,Abd-etal:2013:PRD:,Zah-Fro-Sho:2013:PRD:,Sha-Ata-Ahm:2014:APSS:}. Resonant phenomena were extensively discussed in \cite{Stu-Kot-Tor:2013:AA:}.
 It is well known that small perturbations of circular orbits from equilibrium positions lead to the appearance of quasi-harmonic oscillations. However such oscillations will decay in relatively short timescales due to synchrotron radiation \citep{Shoom:2015:PHYSR4:,Tur-etal:2018:APJ:}. }
This implies that particle's energy at the circular orbit under influence of the radiation reaction force is always located at the minimum of the effective potential.
The radiation reaction force in the locally geodesic frame of reference is aligned against the direction of the motion of the charged particle, concentrating along the motion at a narrow cone. This statement implies that the radiation reaction in fact reduces the velocity of a particle and corresponding kinetic energy. On the other hand, if the motion is stable due to balance of the gravitational and Lorentz forces, reducing the angular velocity of the particle causes the widening of its circular orbit. Such widening leads to increase of particle's energy and angular momentum measured by observer at rest at infinity, while the energy of the particle in the particle's frame always decreases. Below we examine the orbital widening effect for magnetic fields of uniform and dipole configurations.

\subsection{Orbital widening in an asymptotically homogeneous magnetic field}

\begin{figure*}
\includegraphics[width=\hsize]{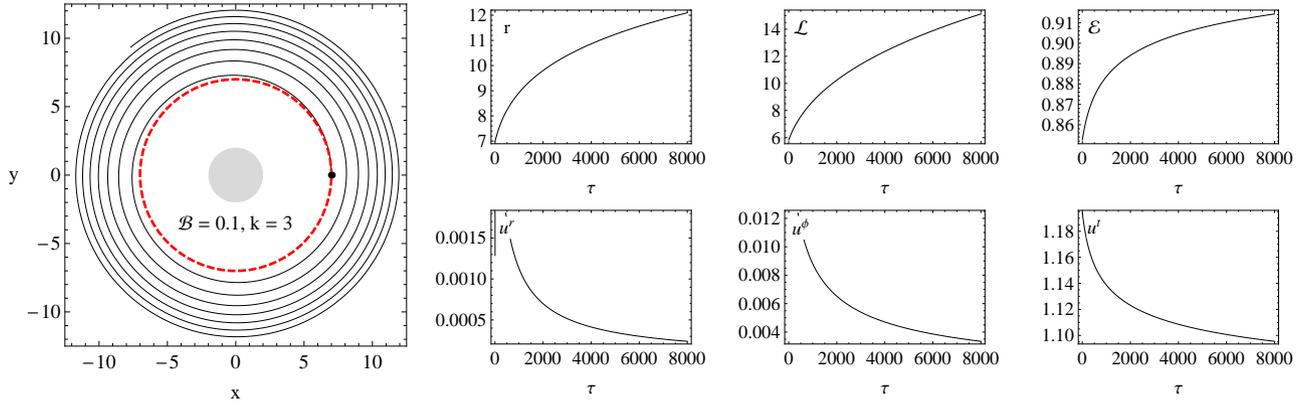}
\caption{\label{trajrad5}
Radiative widening of circular orbit of a charged particle around black hole in uniform magnetic field and corresponding evolutions of orbital radius, angular momentum, energy and different components of velocity of the particle. Starting point on the trajectory is indicated by black dot. The trajectory without radiation is shown by dashed red circle.
}
\end{figure*}

\begin{figure*}
\includegraphics[width=\hsize]{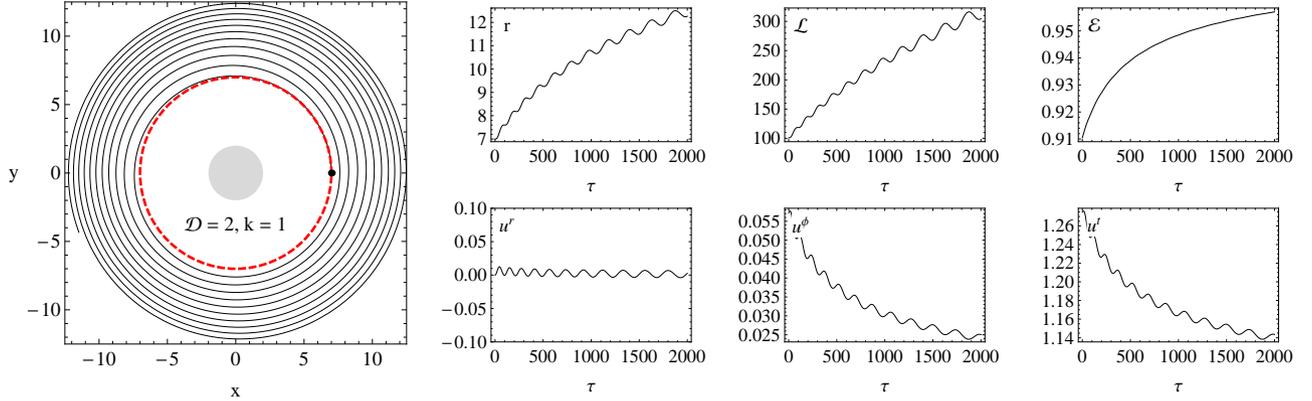}
\caption{\label{wid-dipole}
Same as in Fig.\ref{trajrad5} for magnetic dipole field case.
}
\end{figure*}

Let the black hole is immersed into an asymptotically homogeneous magnetic field which has the strength $B$ at spatial infinity. The field is aligned perpendicularly to the equatorial plane coinciding with the plane of the motion of the charged particles. The corresponding solution of Maxwell's equations in the Schwarzschild spacetime implies the existence of only nonzero covariant component of four-vector potential $A^\mu$ \citep{Wald:1974:PHYSR4:}
\beq
A_{\phi}^{\rm U} = \frac{B}{2} \, r^2 \sin^2\theta. \label{aasbx}
\eeq
Dynamics and corresponding oscillatory motion of charged particles in the absence of radiation reaction force in the magnetized Schwarzschild spacetime can be found in \cite{Kol-Stu-Tur:2015:CQG:} and magnetized Kerr spacetime in \cite{Stu-Kol:2016:EPJC:,Kol-Tur-Stu:2017:EPJC:}. Let us introduce two dimensionless parameters characterizing the influences of magnetic field and radiation reaction force in the black hole vicinity as follows
\beq
{\cal B} = \frac{q B M}{2 m} , \quad k = \frac{2\,q^2}{3 m  M},
\eeq
where $q$ and $m$ are charge and mass of a test particle and $M$ is the black hole mass.

The rate of the energy loss of a charged particle can be calculated directly from the time component of the equation of motion (\ref{curradforce}). Skipping over details, one can write \citep{Tur-etal:2018:APJ:}
\beq \label{enerlossSCH-full}
\frac{d {\cal E} }{d\tau } = - {\cal K}_{1} \ce^3 + {\cal K}_{2} \ce x(\tau),
\eeq
where ${\cal K}_1 = 4 k \BB^2$ and ${\cal K}_2 = 2 k \BB$ are constants and
$x(\tau) = 2 k \BB f (\tau) + u^{\phi} (\tau) / r(\tau)$. The analytical solution of equation (\ref{enerlossSCH-full}) can be found in the form
\beq \label{wideningenergy1}
\ce (\tau) = \frac{\ce_i e^{{\cal K}_2 X(\tau)}}{\left(1 + 2 {\cal K}_1 \ce_i^2 \int_0^{\tau} e^{2 {\cal K}_2 X(\tau')} d\tau'\right)^{\frac12} },
\eeq
where $X(\tau) = \int_0^\tau x(\tau) d\tau$.%
We solve the equation (\ref{enerlossSCH-full}) numerically and illustrate the results in representative plots for the particular %
set of initial conditions.

When the orbital radius $r$ increases faster than the deceleration of velocities $u^t$ and $u^{\phi}$, the energy $\ce$ and angular momentum $\cl$ of the charged particle increase accordingly. Example of the orbital widening of radiating charged particle around Schwarzschild black hole in uniform %
magnetic field is illustrated in Fig.%
\ref{trajrad5}. 
%
 Corresponding changes of the orbital radii in time are given in the $r - \tau$ plot.  If the external Lorentz force is attractive, which corresponds to (${\cal L}\cdot \BB<0$), the charged particle collapses to the black hole. 
In the opposite case, the particle slows down due to radiation reaction while keeping the circular character of the motion. Decrease of particle's velocity $u^\phi$ in stable motion shifts the particle outwards from the black hole. %
 The ${\cal E} - \tau$ plot shows the corresponding increase of particle's energy measured by an observer at rest at infinity, which is in fact the potential energy of the particle, while kinetic energy given by $m \gamma = m u^t$ decreases as the particle slows down due to radiation, as represented in the $u^t - \tau$ plot. The energy of the particle asymptotically tends to its rest energy ${\cal E}_{|\tau \rightarrow \infty} = 1$. If the radiation reaction can be neglected, the trajectory of a particle is circular as indicated by the red dashed circle. Note that the orbital widening can be observed when the energy of a charged particle ${\cal E}<1$, i.e. when the motion of the particle is bounded in the vicinity of a black hole. For ultrarelativistic particles with ${\cal E} \gg 1$ the leading contribution to the evolution of energy is given by the first term on the right hand side of Eq.(\ref{enerlossSCH-full}), leading to the loss of particle's energy.

\subsection{Orbital widening in a magnetic dipole field}

Dipole magnetic field can be generated by circular current loop with radius $R\geq 2 M$, located on the surface of a compact object in the equatorial plane. Outer solution for the electromagnetic 4-vector potential $A^\mu$ in the Schwarzschild metric is given by only nonzero covariant component  \citep{Petterson:1974:PHYSR4:}
\beq
A_{\phi}^{\rm D} = - \frac{3}{8} \frac{\mu r^2 \sin^2\theta}{M^3} \left[ \ln\left( 1 - \frac{2M}{r} \right) + \frac{2M}{r} \left( 1+ \frac{M}{r} \right)  \right] \, , \label{aasby}
\eeq
where $\mu=\pi R^2 (1-2M/R)^{1/2} I$ is the magnetic dipole moment and $I$ is electric current of the loop.
The term in square brackets is negative for $r>2M$. One can parametrize the dynamical equations of charged particles by introducing the relative influence of the dipole magnetic field and the gravitational force as follows
\beq 
{\cal D} = \frac{3 |q| \mu G}{8 m M^2 c^4}.
\eeq
As in the case of the uniform field, for the motion at the equatorial plane one can distinguish two different situations depending on the orientation of the Lorentz force. The widening of the circular orbit occurs only when the Lorentz force is repulsive, i.e. for chosen parametrization, the condition reads ${\cal L \cdot D}>0$,  thus the angular momentum and the magnetic dipole parameters have the same sign. The representative example of orbital widening in the magnetic dipole field is plotted in Fig.\ref{wid-dipole}. In the absence of radiation reaction the orbit is purely circular, as in case of uniform magnetic field. The distinguishing signature of orbital widening in magnetic dipole case is the presence of radial oscillations due to radiation reaction. Anisotropy of the field lines increases the non-linearity of dynamical equations which in the magnetic dipole case can lead to the appearance of quasi-harmonic oscillations of charged particles. The radiation reaction in this case plays a role of small perturbation to the orbit and appearance of a perihelion shift.

\section{Conclusions}\label{sec5}

We have shown that the radiation reaction force can shift the circular orbits  %
 of charged particles outwards from the black hole. This effect occurs only in the case with repulsive Lorentz force, while for opposite orientation the particle spirals down to the black hole. If the stability of the circular %
 orbit is conserved, the radiation reaction acting opposite to the motion of the particle decreases its linear velocity. This in turn leads to the orbital widening which we demonstrated in plots. 
Actual escape of the charged particle from the black hole in the equatorial plane fixed by magnetic field implies that the potential energy of the charged particle increases while the kinetic energy always decreases. Let us estimate the formal maximal efficiency of the gain of potential energy due to orbital widening for the shift of the circular orbit from ISCO to infinity. Defined as the ratio between gained energy to the final energy at infinity ($E_{\infty} = m c^2$), the efficiency depends on the position of ISCO, which is a function of the magnetic field strength for the charged particles \citep{Kol-Stu-Tur:2015:CQG:}. Since the ISCO of a charged particle can be located very close to the event horizon, the maximal efficiency can reach the values up to $100\%$. However, $100\%$ efficiency is unreachable, since it would require the ISCO to be located at the black hole horizon and $\BB, {\cal D} \rightarrow \infty$. 
In the absence of magnetic fields, the formal efficiency of the inverse mechanical process of shifting the particle orbit from infinity to the ISCO is $5.7\%$. One needs to note that in realistic scenarios the effect of radiative widening of orbits is many orders of magnitude slower than the orbital timescales. 
However, in some astrophysical scenarios with large magnetic fields this effect can be potentially relevant.

{Orbital widening accompanied by the increase of the potential energy of charged particle is governed by the second term on the right hand side of Eq.(\ref{enerlossSCH-full}), in the case of uniform magnetic field, while the first term governs the decay of the particle's oscillations around stable circular orbit. When the decay of oscillations is irrelevant, i.e., when the particle is located at the stable circular orbit, one can find the timescale of the orbital widening in the form
\beq \label{tau-wid}
\tau_{\rm W} \approx \frac{1}{2 k \BB} \, {\rm ln}\left|\frac{E}{E_0}\right|,
\eeq
where $E_0$ and $E$ are the initial and final energies of the charged particle. Note that all quantities in Eq.(\ref{tau-wid}) are dimensionless. Inserting back the physical constants we get 
\beq
\tau_{\rm W} = \frac{3 m^2 M G c}{4 q^3 B} \, {\rm ln}\left|\frac{E}{E_0}\right|,
\eeq
where all quantities are given in Gaussian units. For an electron orbiting a stellar mass black hole one can estimate the widening timescale as follows
\beq
\tau_{\rm W} \approx 10^3  \left(\frac{q}{e}\right)^{-3} \left(\frac{m}{m_e}\right)^{2} \left(\frac{M}{10 M_{\odot}}\right) \left(\frac{B}{10^8 {\rm G}}\right)^{-1} \, {\rm s}.
\eeq
Here we assumed ${\rm ln}\left|{E}/{E_0}\right| \sim 1$, which is well justified by previous studies, see, e.g. \cite{Kol-Stu-Tur:2015:CQG:}. One can conclude that the effect of the orbital widening can be relevant in stellar mass black holes and relatively weak in supermassive black holes. More precise calculation of widening timescales can be done by numerical integration of the equations of motion for a particular set of initial conditions.
}

{Another interesting consequence of the orbital widening caused by the radiation reaction force is the appearance of the quasi-harmonic oscillations of charged particle which is a special signature of the presence of a dipole magnetic field. One of the important extensions of our work could be comparison of the frequencies of such oscillations in both radial and vertical modes with mysterious frequencies of the quasi-periodic oscillations observed in many miqroquasars containing black holes or neutron stars.}


\section*{Acknowledgments}


A.T. and M.K. acknowledge the \fundingAgency{Czech Science Foundation} Grant No. \fundingNumber{16-03564Y} and the \fundingAgency{Silesian University in Opava} Grant No. \fundingNumber{SGS/14/2016}. Z.S. acknowledges the \fundingAgency{Albert Einstein Centre for Gravitation and Astrophysics} supported by the \fundingAgency{Czech Science Foundation} Grant No. \fundingNumber{14-37086G}.







\bibliography{Wiley-ASNA}%





\end{document}